\documentclass[review,5p,times,twocolumn]{elsarticle}

\usepackage[dvipsnames]{xcolor}  
\colorlet{greycolor}{red!40!green!40!blue!40!}  
\usepackage{lineno,hyperref}

\usepackage{algorithmicx}
\usepackage[ruled]{algorithm}
\usepackage[noend]{algpseudocode}

\modulolinenumbers[5]

\journal{Journal of Biomedical Informatics}

\usepackage{booktabs}
\usepackage{graphicx}
\usepackage{subfig}
\begin{document}

\begin{frontmatter}

\title{Visual analytics of COVID-19 dissemination in São Paulo state, Brazil}

\author{Wilson E. Marcílio-Jr*, Danilo M. Eler, Rogério E. Garcia, Ronaldo C. M. Correia, Rafael M. B. Rodrigues
}
\address{Department of Mathematics and Computer Science\\
São Paulo State University (UNESP)\\
Presidente Prudente - SP, Brazil
}


\cortext[corr]{wilson.marcilio@unesp.br}


\begin{abstract}
Visual analytics techniques are useful tools to support decision-making and cope with increasing data, which is particularly important when monitoring natural or artificial phenomena. When monitoring disease progression, visual analytics approaches help decision-makers choose to understand or even prevent dissemination paths. In this paper, we propose a new visual analytics tool for monitoring COVID-19 dissemination. We use k-nearest neighbors of cities to mimic neighboring cities and analyze COVID-19 dissemination based on the comparison of a city under consideration and its neighborhood. Moreover, such analysis is performed based on periods, which facilitates the assessment of isolation policies. We validate our tool by analyzing the progression of COVID-19 in neighboring cities of São Paulo state, Brazil.
\end{abstract}


\end{frontmatter}


\section{Introduction}

The novel coronavirus (SARS-CoV-2), or simply COVID-19, has already infected more than 116 million people worldwide and caused over one million deaths by March 2021. While understanding the biological aspects of such a virus is essential~\cite{Huang2020, Qi2020, Xu2020}, it is necessary to monitor the evolution in the number of cases in cities and their neighborhoods to provide information to decision-makers to think about strategies isolation policies according to the risk of dissemination. Moreover, citizens must be aware of how the isolation policies affect the dissemination by COVID-19 and the number of cases after isolation.

To monitor COVID-19 dissemination, we propose a visual analytics tool to help analyze the growth in the number of confirmed cases in the cities of Sao Paulo state, Brazil. ˜ The main contribution of our approach consists of its ability to perform local and regional analysis of the cities by inspecting periods of analysis defined in days. We argue that analyzing a period allows one to follow the evolution of the number of confirmed cases more easily since one can observe the number of cases accumulated for a period and not since the first day of notification. Besides defining a period, our perspective about the regional analysis (a city neighborhood) could be fundamental for decision-making due to disease dissemination patterns that usually follow a hierarchy spreading from bigger cities to neighborhoods. So, besides presenting the number of city confirmed cases, we also present the number of cases in its neighborhood. The comparison between a city and its neighborhood allows verifying if a neighborhood has the main focus of COVID-19 dissemination or various dissemination points. For example, if a city under consideration has a bigger number of cases than its neighborhood, it stands out in the influence of the neighborhood; on the other hand, if the neighborhood stands out over the city under consideration, the neighborhood can have one or more cities that could influence cities with fewer confirmed cases. Finally, if both the city in analysis and its neighborhood have a high number of confirmed cases, such a neighborhood has high COVID-19 dissemination.

To validate our methodology, we provide several case studies by analyzing cities in the Sao Paulo state, Brazil. Besides highlighting the cities on the map, our tool also summarizes the risk of dissemination using a radial visualization. We use the slope of the number of cases to interpret the risk of dissemination. Note that we are focusing on the dissemination risk rather than the number of cases itself. In the radial visualization, the circle encodes the city in the analysis, while a donut chart maps the neighboring cities. We use color saturation to indicate the risk of dissemination. That is, darker colors will represent cities with a higher risk of dissemination. 

This paper is organized as follows: in Section~\ref{sec:related-works}, we briefly delineate some related works; Section~\ref{sec:background} presents the hierarchical spreading of COVID-19, from which our methodology is based; Section~\ref{sec:visualization-design} shows the proposed visual analytics tool; analyses using the tool are presented in Section~\ref{sec:results}; in Section~\ref{sec:discussion}, we discuss some aspects of the technique; we conclude our work in Section~\ref{sec:conclusion}.

\section{Related Works}
\label{sec:related-works}

Using data to detect and quantify health events is a useful strategy to understand disease outbreaks. Usually, the strategies use data mining or visualization techniques to monitor events related to a disease of interest. 

Visualization-based strategies for monitoring the dissemination of diseases account for the fact that graphical representations can enhance the ability to identify data patterns and tendencies. In this case, it is better to look at visual variables, such as position, color, or area, than tables or reports to identify tendencies of growth and many other patterns. The literature presents many examples of systems using visualization techniques to enhance the analysis of disease dissemination, such as the work of Hafen et al.~\cite{Hafen2009}, where they delineate a strategy to detect outbreaks based on monitoring pre-diagnostic data of emergency department chief complaints. Although using simple line curves, the visualization components helped at identifying patterns in the data. HealthMap~\cite{Freifeld2008}, on the other hand, uses geolocation of media reports to integrate outbreak textual data in a single resource. The system helps extract useful information and summarize unstructured data of disease reports, facilitating decision-makers analysis. Another interesting approach is to use what-if strategies and visualize the outcomes depending on the decision alternatives applied when dealing with disease outbreaks~\cite{Brigantic2010}. Other strategies employing visualization tools are using heatmaps to analyze patterns of hand-foot-mouth disease~\cite{Hand2015}, employing intelligent graph visualizations and reordered matrices to understand influenza dissemination paths~\cite{Guo2007}, or visualizing the effect of decision measures implemented during a simulated pandemic influenza scenario~\cite{Maciejewski2011}.

From the data mining perspective, it is usually interesting to contrast social network posts related to diseases with officially reported cases. These approaches are based on the strength of the relationship between officially reported cases and the searches on the web or posts on social media using words related to the diseases~\cite{Paul2014, Culotta2010, Achrekar2011, Broniatowski2013}. For instance, most of the works use the web and social media to detect Influenza-like Illness events~\cite{Paul2014, Santillana2015, Astill2018}. An excellent example of using data-mining techniques to detect disease outbreaks in the dutch system Coosto~\cite{Belt2018}, which uses Google Trends and social media data to detect outbreaks using a cut-off criterion. Other works also have shown that Twitter data is highly correlated to disease activity~\cite{Gomide2011,Coelho2016}, such as predicting Dengue cases~\cite{MarquesToledo2017} or using Twitter-based data to automatically monitor avian influenza outbreaks, showing that one-third of outbreak notifications were reported on Twitter earlier than official reports~\cite{Yousefinaghani2019}. 

In this work, we provide a visual analytics approach to monitor the evolution of dissemination of COVID-19 to facilitate analysis of dissemination during and post isolation. We use visualization techniques and analysis based on time windows to help analysts monitor how neighboring cities' situation affects the dissemination of COVID-19 to a city in the analysis.

\section{Background}
\label{sec:background}

This section explains the hierarchical spreading behavior of COVID-19, which we use to define each city's neighborhoods.

\begin{figure}[!htb]
  \centering
  \includegraphics[scale=0.8]{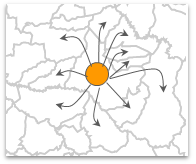}
  \caption{The neighborhood of a city (represented by an orange circle) consists of the nearest cities according to the distances computed by latitude and longitude coordinates.}
  \label{fig:hierarchical-spreading}
\end{figure}

The basic idea of the hierarchical spreading of COVID-19 is that cities with confirmed cases disseminate infection to their neighboring cities. In this case, the neighboring cities are the cities with confirmed cases inside the $k$ nearest cities to the city in the analysis. Note that regional cities (bigger cities) are most likely to disseminate COVID-19 to their neighborhoods due to the bigger number of inhabitants, more job opportunities, more cultural access, and other social aspects that could attract people from neighboring cities. Figure~\ref{fig:hierarchical-spreading} illustrates the hierarchical spreading, where an orange circle represents a city with a confirmed case, and the arrows indicate potential dissemination paths due to location proximity.

To formalize the definition of a neighborhood for city A, Algorithm~\ref{alg:nearest-cities} shows the computation of each city's $k$ nearest cities in the Sao Paulo state based on the latitude and longitude coordinates.

\begin{algorithm}[]
\small
\caption{Computing $k$ nearest cities.}
\label{alg:nearest-cities}
\begin{algorithmic}[1]
\Procedure{k\_nearest\_cities}{cities, $k$}
    \State latlong\_coords $\leftarrow$ get\_latlong(cities);
    \State knn\_sets $\leftarrow$ KNN(latlong\_coords, $k$);
    \State \Return knn\_sets;
\EndProcedure
\end{algorithmic}
\end{algorithm}

To augment city $A$'s neighborhood, besides the $k$ nearest cities of $A$, we also add to the neighborhood the cities that have $A$ in their $k$ nearest cities set. In this way, we simulate better the interaction between a city and its neighborhood. Finally, it is essential to mention that a few cities in the neighborhood of a city of interest do not have confirmed cases. In this case, these cities do not appear in the visualization tool (see Section~\ref{sec:visualization-design}).

\section{Visualization Design}
\label{sec:visualization-design}

Our visual analytics approach has the main objective of helping decision-makers analyze the situation of a city based on disease dissemination. Besides information of a city in interest, our tool provides information about the situation of its neighborhood. We delineate the following requirements for our visual analytics approach to monitor the dissemination curves and help analysis based on the number of infections by COVID-19::

\begin{itemize}
    \item[] \textbf{R1}: facilitate comparison of the situation between a city and its neighborhood;
    \item[] \textbf{R2}: visualize the evolution of the number of cases as users change a time window, as well as contrasting it with the accumulated number of cases since the notification of the first confirmed case;
    \item[] \textbf{R3}: visualize the dissemination curve to check if it is increasing or if it is flattening;
    \item[] \textbf{R4}: quickly understand the situation of a neighborhood in the analysis.
\end{itemize}

First, it is necessary to define the neighborhood of a city under consideration. Our strategy to define the neighborhood follows the hierarchical spreading scheme of COVID-19, as explained in Section~\ref{sec:background}, which states that a city with confirmed cases influence (i.e., can disseminate) its neighboring cities. These neighboring cities are retrieved using the $k$ nearest neighbors algorithm. In our case, city $A$'s neighborhood consists of the union of the $k$ nearest cities to $A$ and the cities with $A$ in their $k$ nearest cities set. Given that, we can analyze a city based on its dissemination as well as its neighborhood. In the following, we present how we accomplish each requirement.

\begin{figure*}[!htb]
  \centering
  \includegraphics[width=\linewidth]{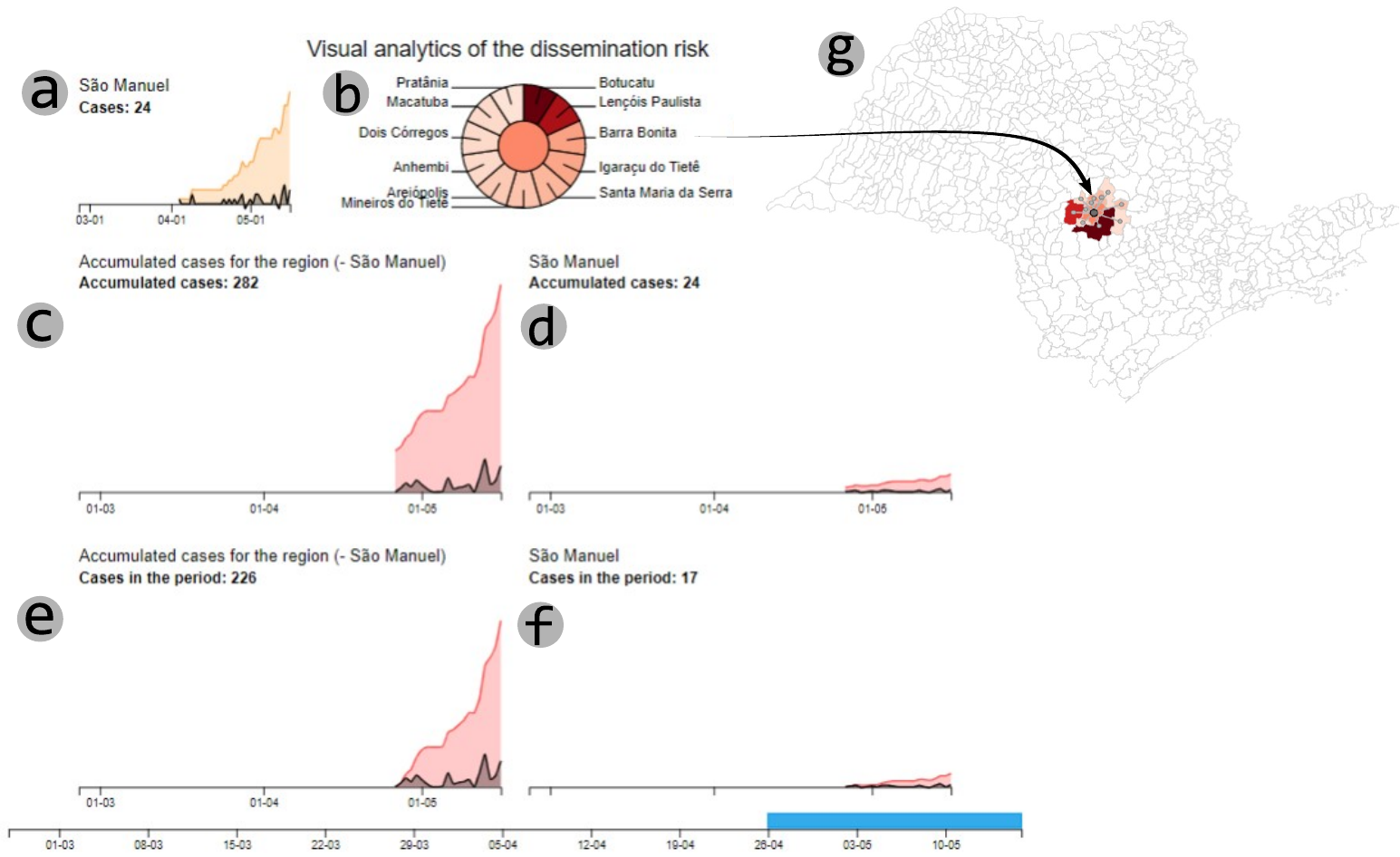}
  \caption{Visual analytics of COVID-19 confirmed cases. (A) The number of cases for the whole period. (B) The visual representation of the analyzed neighborhood with color saturation encoding a high slope of new cases. The accumulated number of cases for a period in the neighborhood (C) and the city in the analysis is based on the total number of cases (D). The accumulated number of cases for a period in the neighborhood (E) and in the city (F) based only on the number of cases inside the time window. (G) São Paulo state map showing the selected city and its neighborhood. }
  \label{fig:ferramenta}
\end{figure*}

Figure~\ref{fig:ferramenta} shows the tool used to monitor the evolution of COVID-19 in the São Paulo state, Brazil, and the dissemination risk based on city neighborhoods. The tool has a few components. First, the evolution of the number of cases for the whole period starting from February 27th (2020) locates at the top left (a). Second, we provide a visual representation of the dissemination risk by summarizing the neighborhood of the analyzed city at the top center of the visualization (b). The color saturation depicts the dissemination risk – darker colors represent cities with more critical situations – mapped from the angle formed by the slope of COVID-19 cases. That is, the color saturation maps the angle below the line segment formed by the point ($a$, $n_a$) and ($b$, $n_b$), where $n_a$ is the number of cases in the first day of the time window ($a$) and $n_b$ is the accumulated number of cases in the period (from $a$ to $b$). Third, we provide the curves of the number of cases in the central area of the component. The first line of curves indicates the number of cases for the whole period since the first notification of confirmed cases ((c) and (d)). The second line of curves indicates the number of cases only inside the time window ((e) and (f)), which could be useful to assess how isolation policies are affecting the dissemination in a period – note that the graph is generated based on the number of cases in the first day of the period ($a$) and the accumulated number of cases for the period (from $a$ to $b$). Finally, the representation of the time window (period of analysis) is shown at the bottom of the visualization. We also provide a map of the São Paulo state (g) to visualize how the neighborhood in the analysis.

\paragraph{\textbf{R1}: Facilitate comparison between city and its neighborhood}

\begin{figure}[!htb]
  \centering
  \includegraphics[width=\linewidth]{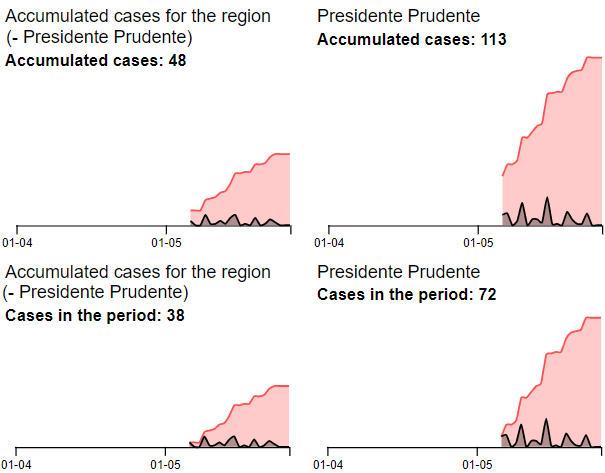}
  \caption{Curves for accumulated number of cases analysis (on top) and analysis based on the time window (on bottom).}
  \label{fig:requirement-r1}
\end{figure}

To facilitate a comparison between a city and its neighborhood, users can use the infection curves for the city itself and the neighborhood. With such visualization, it is possible to understand if the city is being influenced by its neighborhood when the number of confirmed cases is greater for the neighborhood or if the city influences the neighborhood when the number of confirmed cases is greater for the city. Figure~\ref{fig:requirement-r1} illustrates a city influencing its neighborhood in the time window and the total number of cases. Note that the number of cases in Presidente Prudente is by far bigger than in its neighborhood.

\paragraph{\textbf{R2}: Visualize the evolution of number of cases in a time window}

\begin{figure}[!htb]
  \centering
  \includegraphics[width=\linewidth]{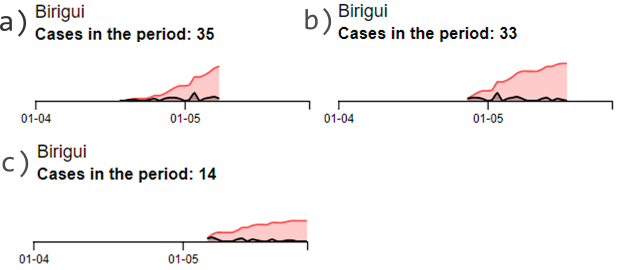}
  \caption{Using a sliding time window allows analyzing the evolution of the number of cases in periods. Such an approach facilitates the inspection of curve patterns.}
  \label{fig:requirement-r2}
\end{figure}

Using a time window, i.e., focusing analysis in a specified number of days, helps monitor the evolution of dissemination in chunks of time and facilitates the comparison among cities. In this case, questions such as which city responds better to isolation policies and how the isolation policy in a certain period affected the dissemination of a posterior period can be answered. Figure~\ref{fig:requirement-r2} illustrates how using the time window for analyzing only the confirmed cases inside the period helps us visualize flattening the Birigui city's curve.

\paragraph{\textbf{R3}: Visualize the increasing and flattening of the curves}

Although the curves with the total number of notification are sufficient to communicate how a city or neighborhood is performed on a period, the curves consisting of only reported cases in a period of analysis help users at focusing only on the increasing or flattening aspects of the curve, as shown for the requirement \textbf{R2} and the example presented in Figure~\ref{fig:requirement-r2}. Such an approach is also useful when contrasting the curve slope with isolation indices in previous periods.

\paragraph{\textbf{R4}: Quickly understanding the situation of a neighborhood}

\begin{figure}[!htb]
  \centering
  \includegraphics[width=0.8\linewidth]{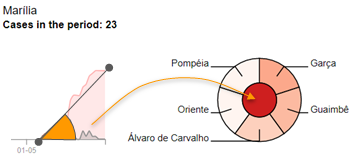}
  \caption{Donut chart encoding neighborhood information. The inner-circle color encodes the angle formed by the city's dissemination curve in analysis, while segment colors encode the angles formed by the dissemination curves of neighboring cities.}
  \label{fig:requirement-r4}
\end{figure}

To promptly visualize a neighborhood's situation, we designed a glyph inspired in Somarakis et al.'s work~\cite{bib:2019_imacyte} to encode the slope of the dissemination curves in a period. A donut chart represents a city's neighborhood in our visualization, while a concentric circle encodes the analyzed city, as shown in Figure~\ref{fig:requirement-r4}. We use color to represent the angle formed by the slope of the dissemination curves. That is, the bigger the increase in the number of cases, the darker the color. Notice that the colors encode the increase, not the number of cases.

\subsection{Data Source}

To monitor the evolution of confirmed cases, we use the São Paulo government's data at SEADE\footnote{https://www.seade.gov.br/coronavirus/}. The data consists of daily updated cases in each city with a confirmed case. To create the visualization tool, we only use the city name, daily confirmed cases, and the date attributes – the latitude and longitude coordinates of each city are retrieved using the \texttt{geopy}\footnote{https://geopy.readthedocs.io/} library. Further, the isolation index is also provided by the São Paulo government\footnote{https://www.saopaulo.sp.gov.br/coronavirus/isolamento/} for only a few cities. The isolation index consists of the percentage of inhabitants in isolation. 

Given the daily updated data, we can use a time window to investigate the progression and dissemination of COVID-19 since the first notification in Sao Paulo state, in the capital São Paulo. Notice that, while using a time window of seven days (without loss of generalization), we could investigate disease progression in the next few days in São Paulo's neighborhood. So, it is not our tool's concern to select cities hypothesizing that they will disseminate the virus. Instead, investigators select a city of interest and then use the time window to visualize and understand the first notification and the progression of confirmed cases compared to its neighbors.

\section{Results}
\label{sec:results}

In this section, we inspect the dissemination evolution of various cities in the Sao Paulo state, Brazil. We used a time window of 20 days to assess the following periods: from April 19th to May 9th, from April 21st to May 11th, and from April 26th to May 16th.

\subsection{Regional city presents risk of dissemination to its neighbors}

\paragraph{Presidente Prudente} 

\begin{figure}[!htb]
  \centering
 \includegraphics[width=\columnwidth]{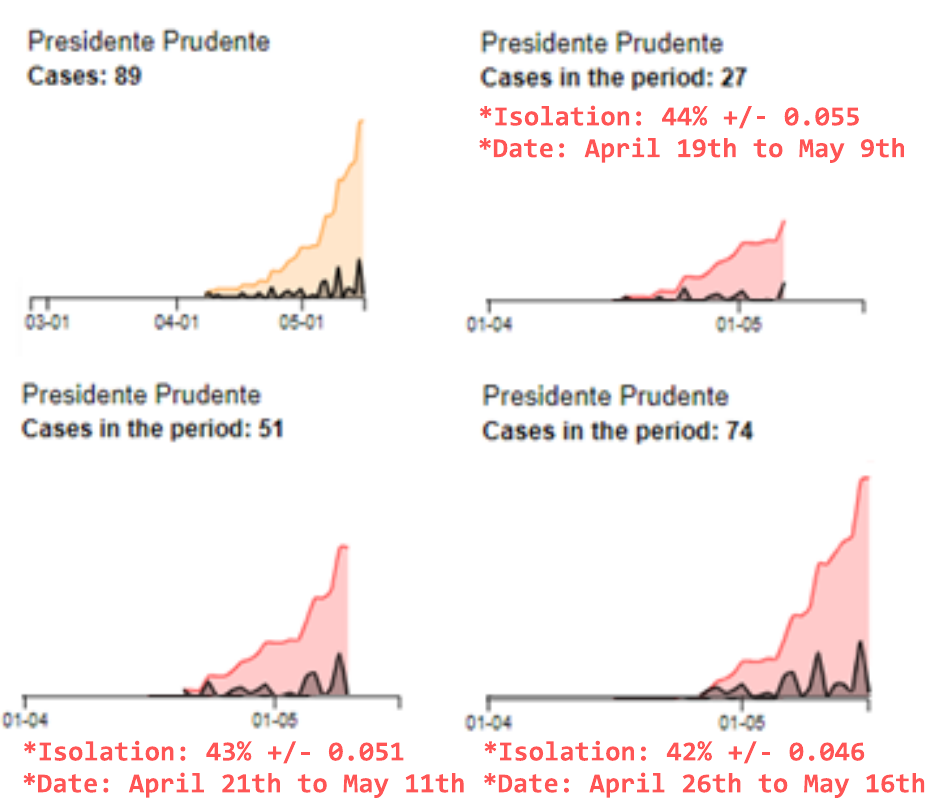}
  \caption{Evolution of the number of cases in three different periods. There is an increasing pattern in the number of cases, which the low isolation indices could explain.}
  \label{fig:PresidentePrudente-analysis}
\end{figure}

Fig.~\ref{fig:PresidentePrudente-analysis} shows the evolution of the confirmed cases of Presidente Prudente. Up to May 16th, Presidente Prudente has 89 confirmed cases, as shown on the figure's left. However, it is interesting to note how was the acceleration of confirmed new cases in the city in the period of April 19th to May 9th, April 21st to May 11th, and April 26th to May 16th. This city, in particular, has been giving the lowest isolation indices in the Sao Paulo state – we highlight in red the mean of the isolation indices in the period.

\begin{figure}[!htb]
  \centering
  \includegraphics[width=\columnwidth]{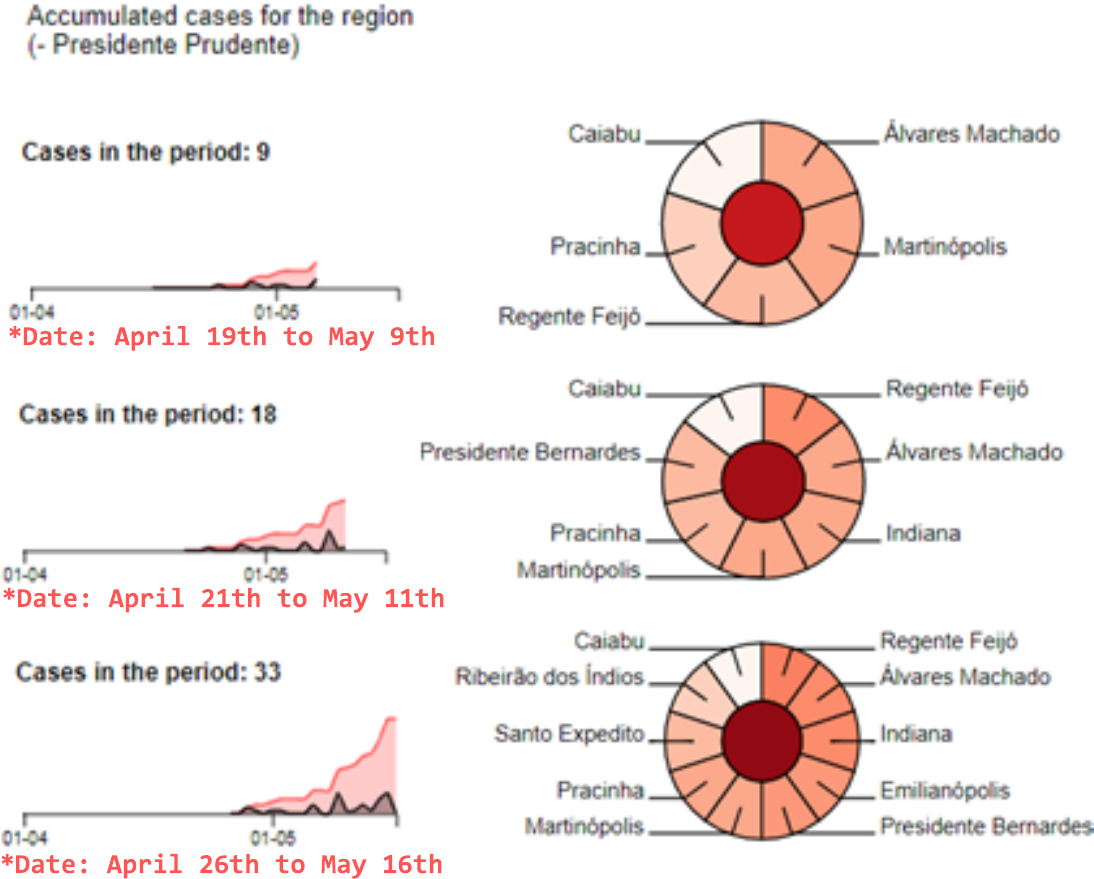}
  \caption{Using the donut chart to assess the evolution of confirmed cases in the neighborhood of Presidente Prudente. Presidente Prudente is the most influential city (a higher number of cases).}
  \label{fig:PresidentePrudente-region}
\end{figure}

It is worth noticing that besides presenting a critical situation in the city itself, Presidente Prudente influence a lot in its neighborhood. Fig.~\ref{fig:PresidentePrudente-region} shows how the number of cases in the city grows as the time window moves (observe the circle in the center of the donut chart), besides the number of different cities presenting confirmed cases. For instance, the increasing number of cases in the neighborhood can be seen by looking at the period's curves. Finally, see how Caiabu maintains the risk of dissemination lower than the other cities through the days.

\paragraph{Martinópolis}

After inspecting Presidente Prudente, we could notice that such a regional city greatly influences its neighborhood according to the risk of dissemination. In this case, given the low isolation index of Presidente Prudente, it could be useful to understand how neighboring cities can respond to such risk. Taking the city of Martinopolis as an example, we can see from Fig.~\ref{fig:PresidentePrudente-region} in the radial representation that social distancing policies might help the city maintain the number of cases. See how in the first period (from April 19th to May 9th), the city was the second in the risk of dissemination, while in the last period of analysis (from April 26th to May 16th), the city lost position to many others.

\begin{figure*}[!htb]
  \centering
  \includegraphics[width=\linewidth]{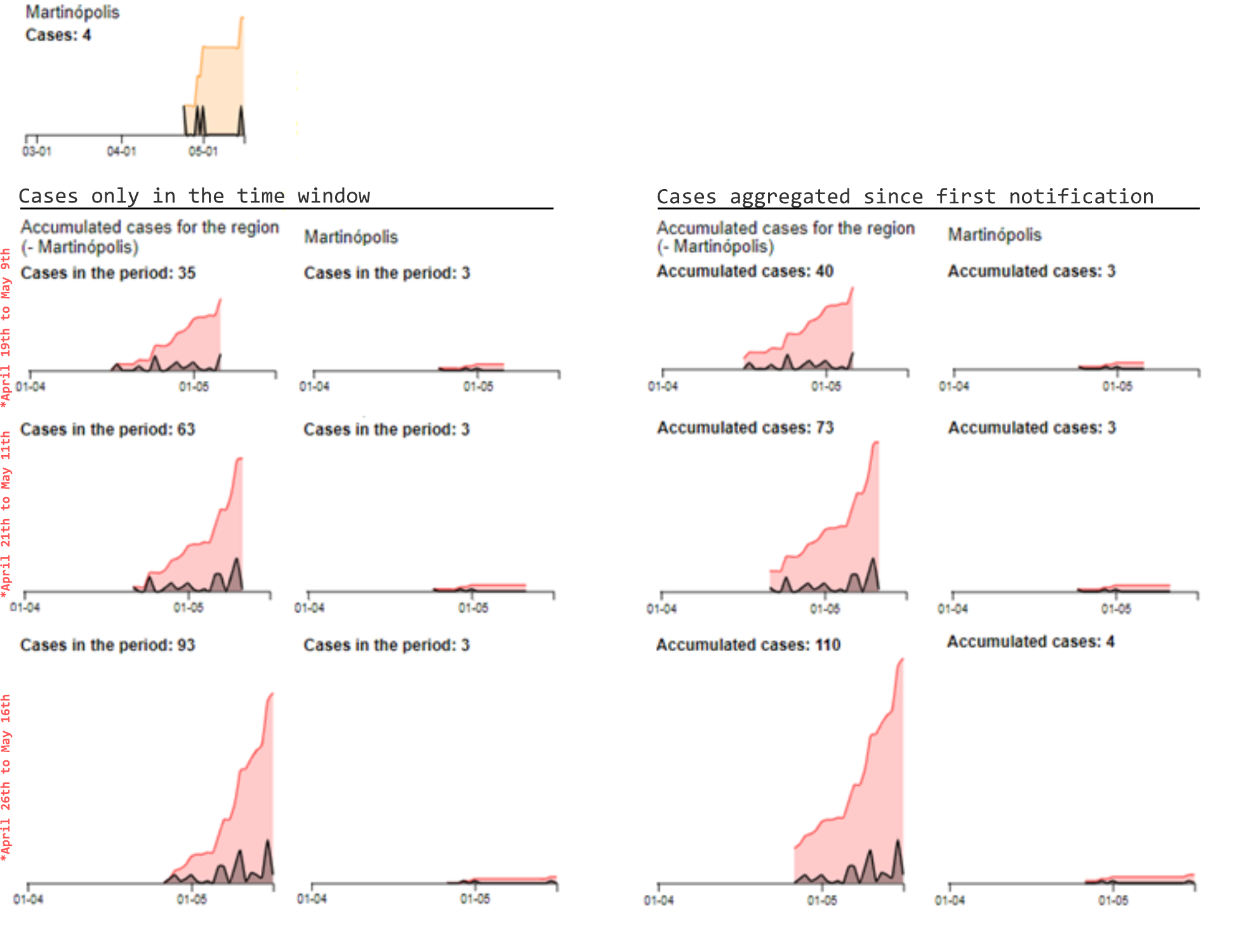}
  \caption{Comparing Martinópolis and its neighborhood in terms of dissemination curve. The city seems to cope well with the increase in cases in the neighborhood.}
  \label{fig:Martinolopis-analysis}
\end{figure*}

Fig.~\ref{fig:Martinolopis-analysis} illustrates the evolution of confirmed cases for the city of Martinopolis. While in the three different period, the number of aggregated cases for the neighborhood grows larger, mainly due to the increase in the number of cases in Presidente Prudente, the number of cases in Martinopolis only increases by one confirmed cases in the period between April 26th to May 16th, reaching only four confirmed cases.

\paragraph{Alfredo Marcondes}

Unlike most of the cities in the neighborhood of Presidente Prudente, Alfredo Marcondes did not present any confirmed cases by the time of analysis. However, being inserted in such a complicated area, the city government could use such information to the aware population of the city's risk.

\begin{figure}[!htb]
  \centering
  \includegraphics[width=0.9\linewidth]{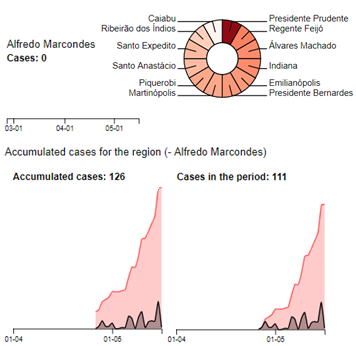}
  \caption{Dissemination curves for the city of Alfredo Marcondes in the period from April 26th to May 16th. In the neighborhood, Presidente Prudente seems to influence others by presenting a high increase in the number of cases.}
  \label{fig:AlfredoMarcondes-analysis}
\end{figure}

\subsection{Neighboring cities with different isolation indices show a different risk of dissemination}

This section analyzes the number of cases and the risk of dissemination by comparing two neighboring cities, Araçatuba and Birigui. Here, we also use these cities' isolation indices to hypothesize about their relation to the number of cases, although we would need statistical analysis to make hard conclusions. 

\begin{figure}[!htb]
  \centering
  \includegraphics[width=\linewidth]{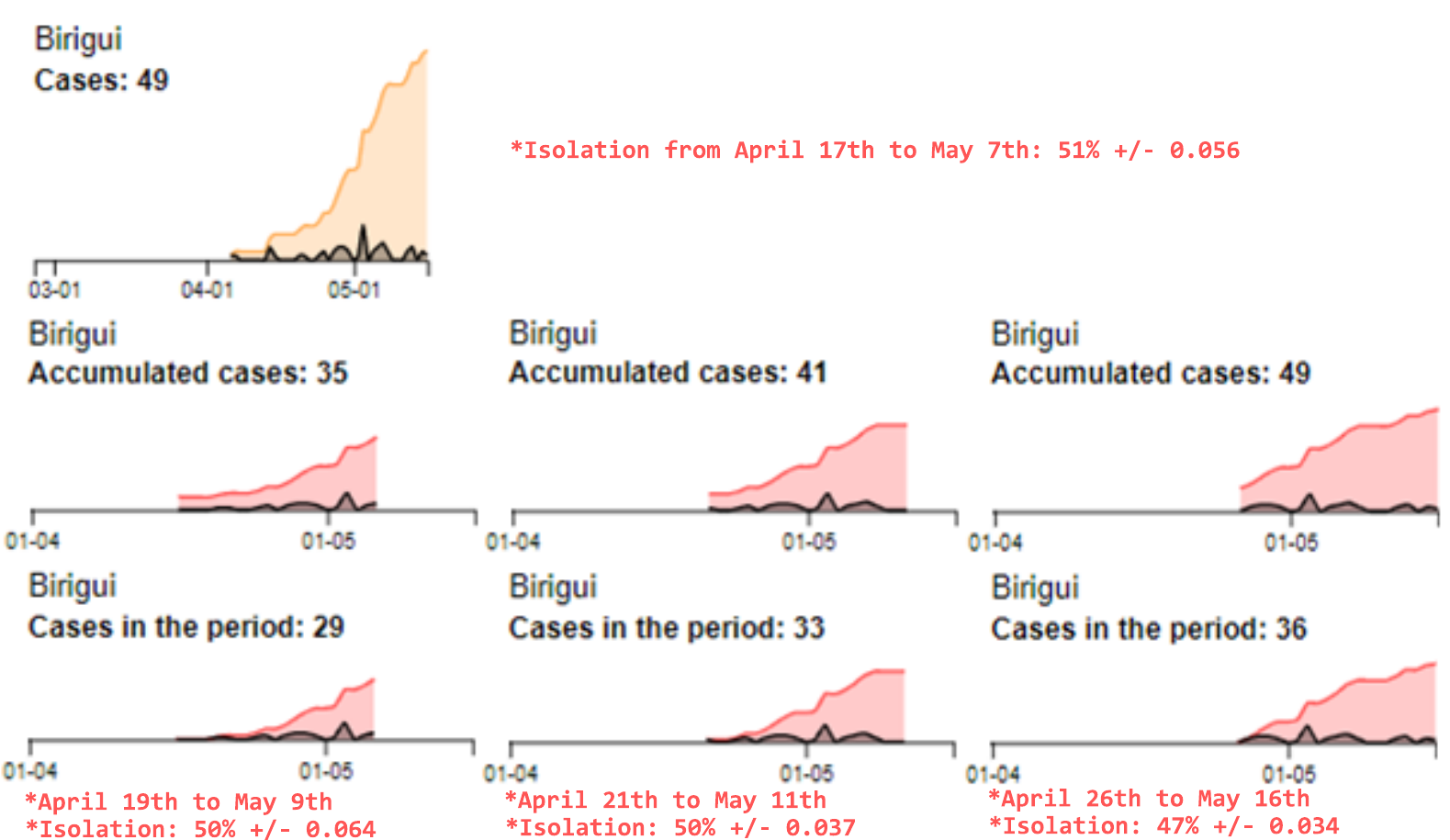}
  \caption{Visualizing the evolution of number of cases in Birigui. By contrasting the curves of cases and the isolation indices, it can be noticed the effect of isolation indices in latter periods of time.}
  \label{fig:Birigui-analysis}
\end{figure}

Fig.~\ref{fig:Birigui-analysis} shows the curves for the city of Birigui. From contrasting the curves in the bottom with the isolation index, we can see how it affects the dissemination of the COVID-19. From the second to the third period, the curve of cases starts to present a plateau. 

\begin{figure}[!htb]
  \centering
  \includegraphics[width=\linewidth]{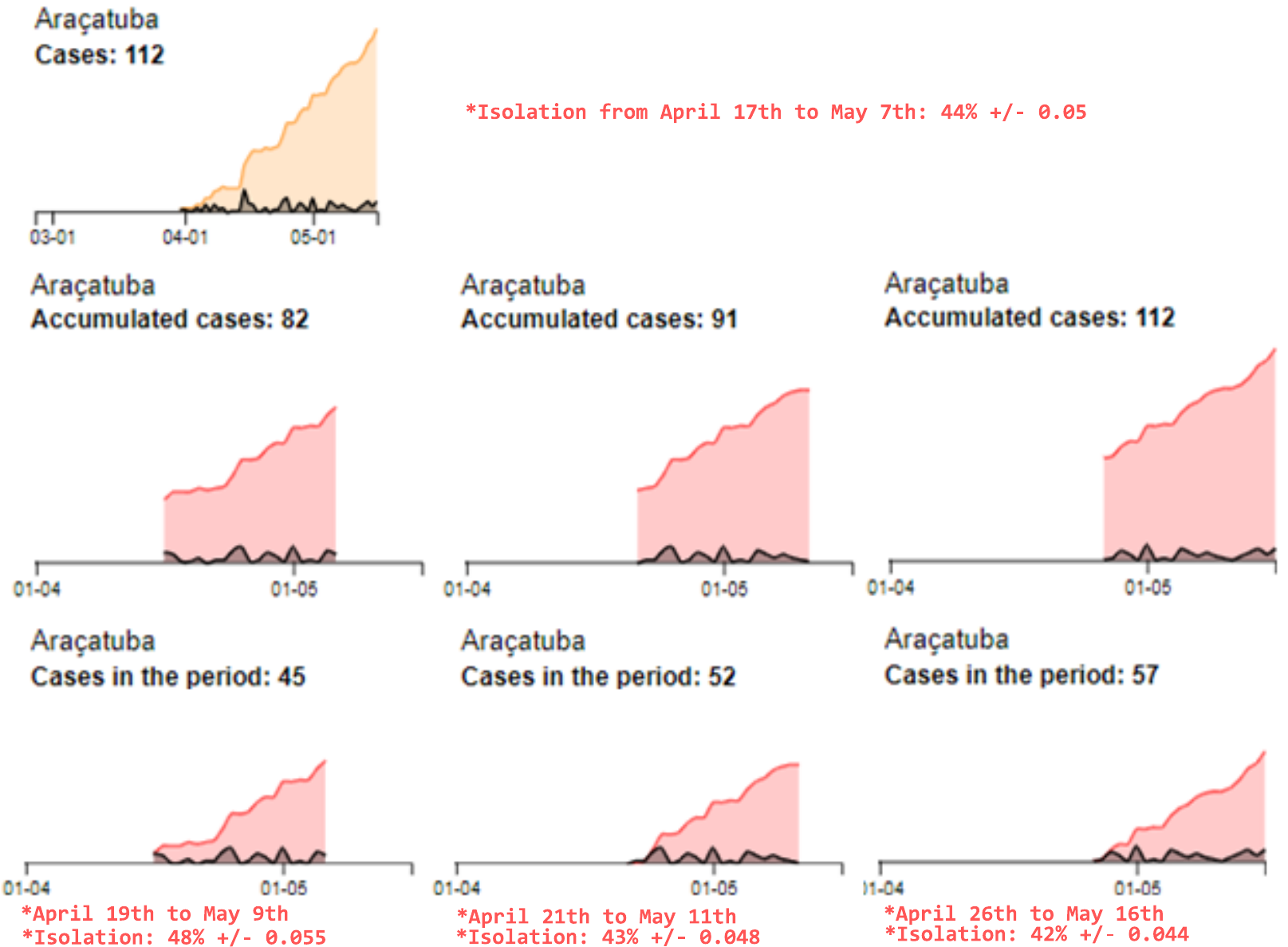}
  \caption{Evolution of the number of cases in Araçatuba. The increasing curve of confirmed cases of Araçatuba could be explained by the low isolation indices in all of the periods of analysis.}
  \label{fig:Aracatuba-analysis}
\end{figure}

In Fig.~\ref{fig:Aracatuba-analysis}, we show the dissemination curves and the isolation indices for Araçatuba. In this case, low isolation indices might be the reason for the increase in the number of cases. Araçatuba has a more critical dissemination curve, which could be explained – together with other reasons – by the low isolation index, which was 44\% +/- 0.05 from April 17th to May 7th. Finally, due to the augment to 48\% +/- 0.055 from April 19th to May 9th, it is possible to see a little flattening in the curve from April 21st to May 11th.

\subsection{The effect of the isolation index}

This section aims to analyze cities in neighborhoods with a rapidly increasing number of cases, besides making relations to isolation indices for hypothesis generation. For this purpose, we analyze three periods: from March 23th to April 12th, from April 13th to May 3rd, and from May 1st to May 21st. 

\begin{figure}[!htb]
  \centering
  \includegraphics[width=\linewidth]{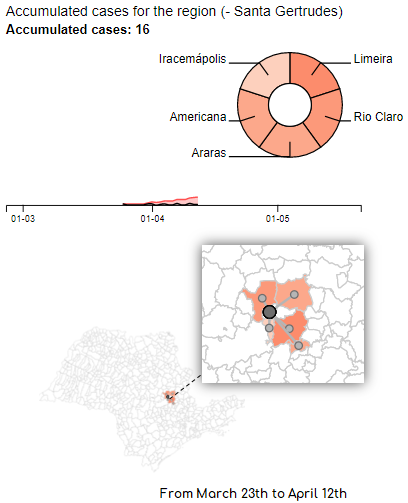}
  \caption{The situation in the city of Santa Gertrudes and its neighborhood from March 23rd to April 12th.}
  \label{fig:SantaGertrudes_first-period}
\end{figure}

Fig.~\ref{fig:SantaGertrudes_first-period} shows how was the situation in the period from March 23th to April 12th. We can see that only five cities presented confirmed cases – 16 confirmed cases in total among all neighboring cities. On the right, it is possible to see how these neighboring cities on the map. 

\begin{figure}[!htb]
  \centering
  \includegraphics[width=\linewidth]{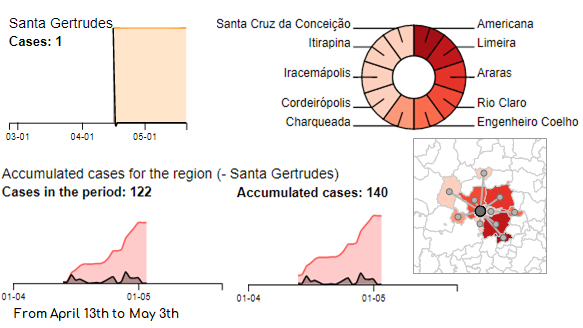}
  \caption{The situation in the city of Santa Gertrudes and its neighborhood from April 13th to May 3rd. The number of confirmed cases and the number of neighboring cities with confirmed cases start to increase.}
  \label{fig:SantaGertrudes_second-period}
\end{figure}

The neighborhoods' situation takes a real change from April 13th to May 3rd, as seen in Fig.~\ref{fig:SantaGertrudes_second-period}. In this case, we see many cities presenting confirmed cases, led by the cities of Americana and Limeira. The city of Americana, for example, was only the penultimate city with the highest risk of dissemination from March 23th to April 12th (see Fig.~\ref{fig:SantaGertrudes_first-period}). For instance, this period was responsible for increasing 122 (of 140) by May 3rd.

\begin{figure}[!htb]
  \centering
  \includegraphics[width=\linewidth]{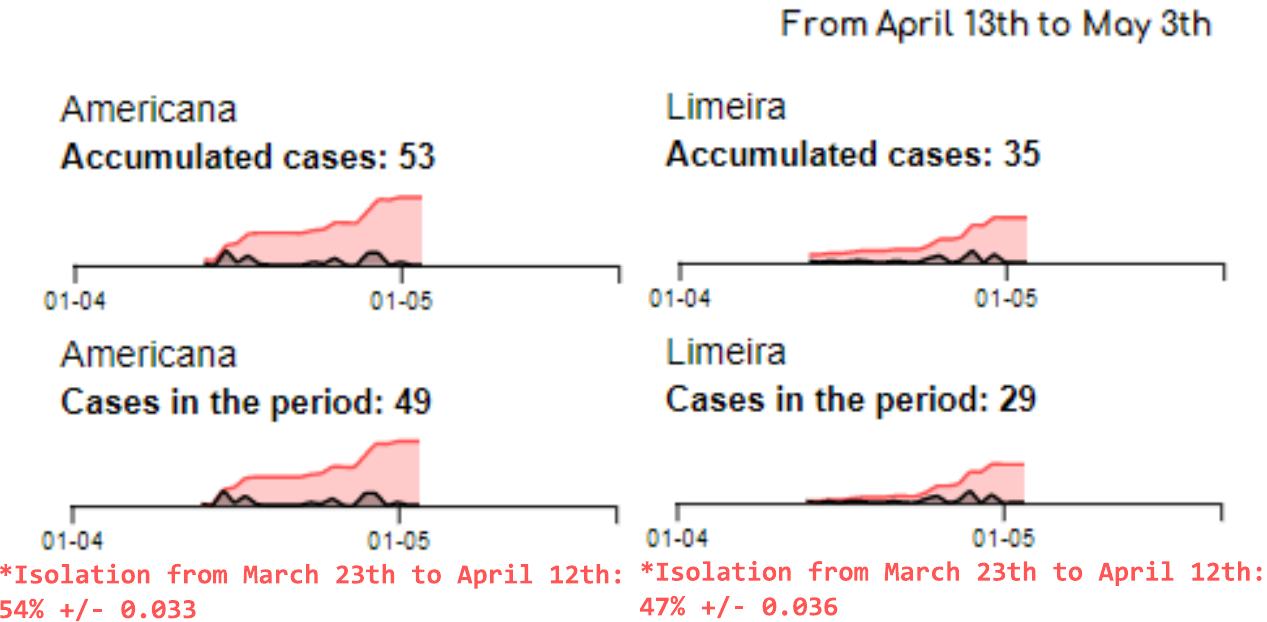}
  \caption{Although Americana presented a higher isolation index for an earlier period, Limeira reported fewer cases from April 13th to May 3rd.}
  \label{fig:AmericanaLimeira-isolation}
\end{figure}

To understand why Americana and Limeira's cities present such an increasing number of confirmed cases, Fig.~\ref{fig:AmericanaLimeira-isolation} shows their curves for the period in the analysis together with the isolation index of an earlier period, i.e., from March 23th to April 12th. In this case, we see that although the isolation index of Americana was greater than Limeira's, the city presented more cases in the period, even with a lower population – Americana has approximately 230 hundred inhabitants while Limeira has approximately 300 hundred inhabitants. The answer to this question can be found by analyzing the neighborhoods of both Limeira and Americana.

\begin{figure}[!htb]
  \centering
  \includegraphics[width=\linewidth]{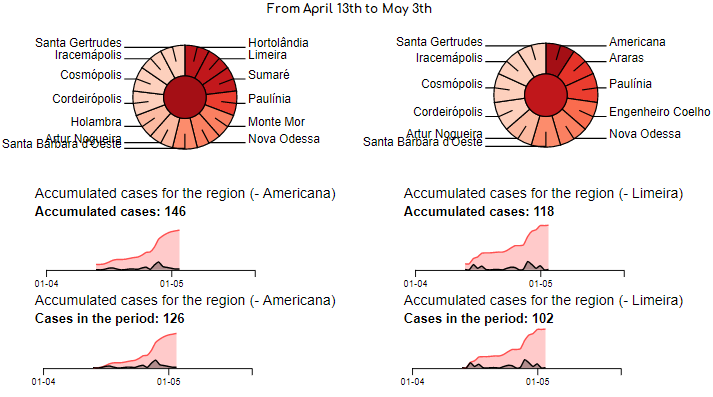}
  \caption{Comparing the neighborhoods of Americana and Limeira. For earlier periods, Americana showed higher isolation indices than Limeira. However, it is part of a critical neighborhood.}
  \label{fig:AmericanaLimeira-neighborhood}
\end{figure}

Fig.~\ref{fig:AmericanaLimeira-neighborhood} explains why Americana shows more number of cases from April 13th to May 3rd. Firstly, the accumulated number of cases in the neighborhood is bigger for Americana. Second, we can notice from the donut chart that Americana is part of a riskier neighborhood, i.e., the number of cities influencing Americana is bigger and the risk of contamination of these cities (darker colors).

\begin{figure}[!htb]
  \centering
  \includegraphics[width=\linewidth]{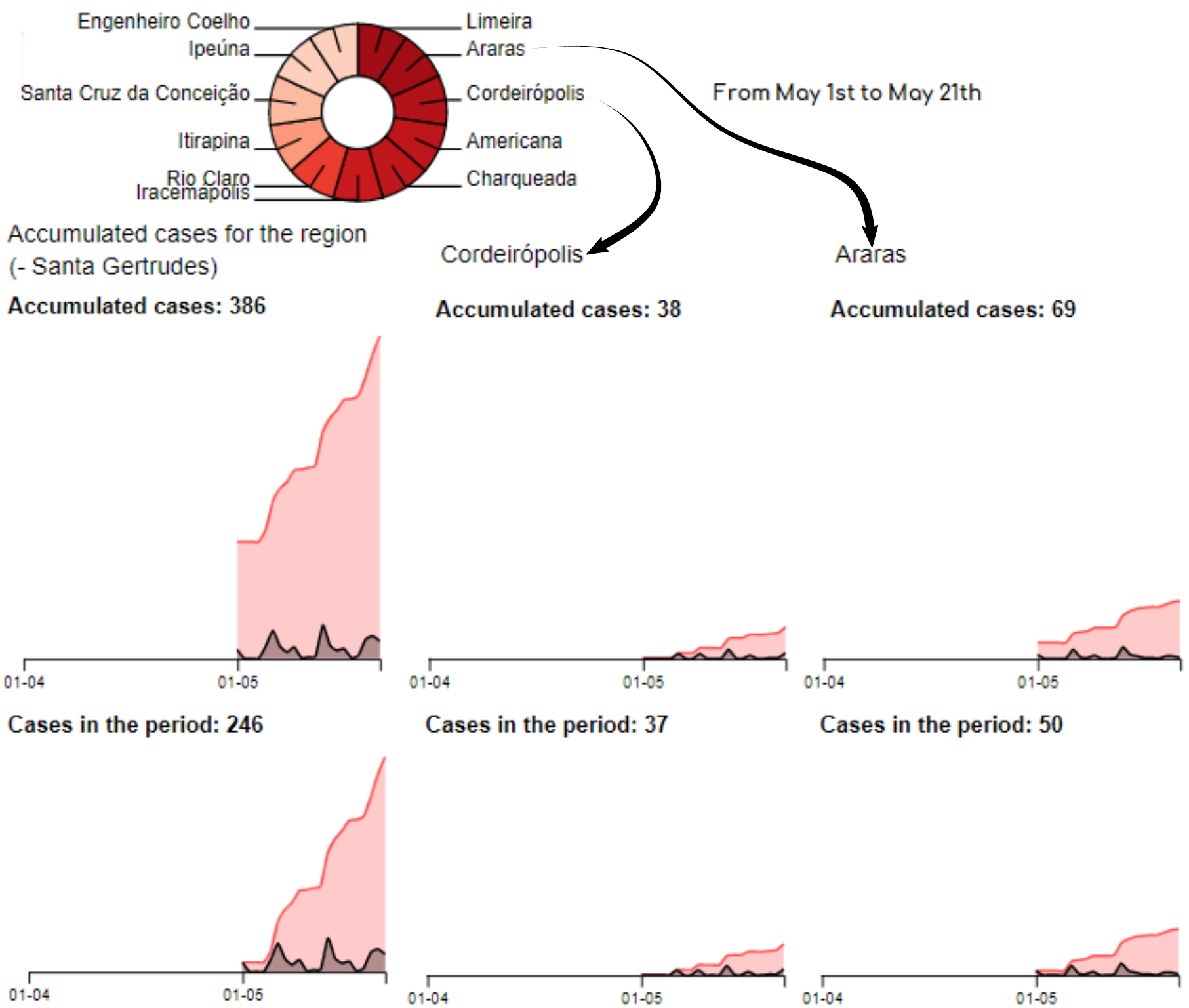}
  \caption{Cities in the neighborhood of Santa Gertrudes showing increase patterns in the number of confirmed cases. New cities start to present a rapid increase in the third period of analysis (from May 1st to May 21st).}
  \label{fig:SantaGertrudes_third-period}
\end{figure}

Backing to the Santa Gertrudes’s neighborhood, we finish by analyzing the last period, from May 1st to May 21st. Fig.~\ref{fig:SantaGertrudes_third-period} shows the situation of the neighborhood up to May 21st. The first thing to notice is the curve inclination of the cases in the period, where it is possible to see the exponential pattern of COVID-19. Then, we can see that two other cities (Araras and Cordenopolis) notified almost the total number of cases in this period. The influence that the risk of the neighborhood plays in these cities can explain such a pattern. 

\begin{figure}[!htb]
  \centering
  \includegraphics[width=\linewidth]{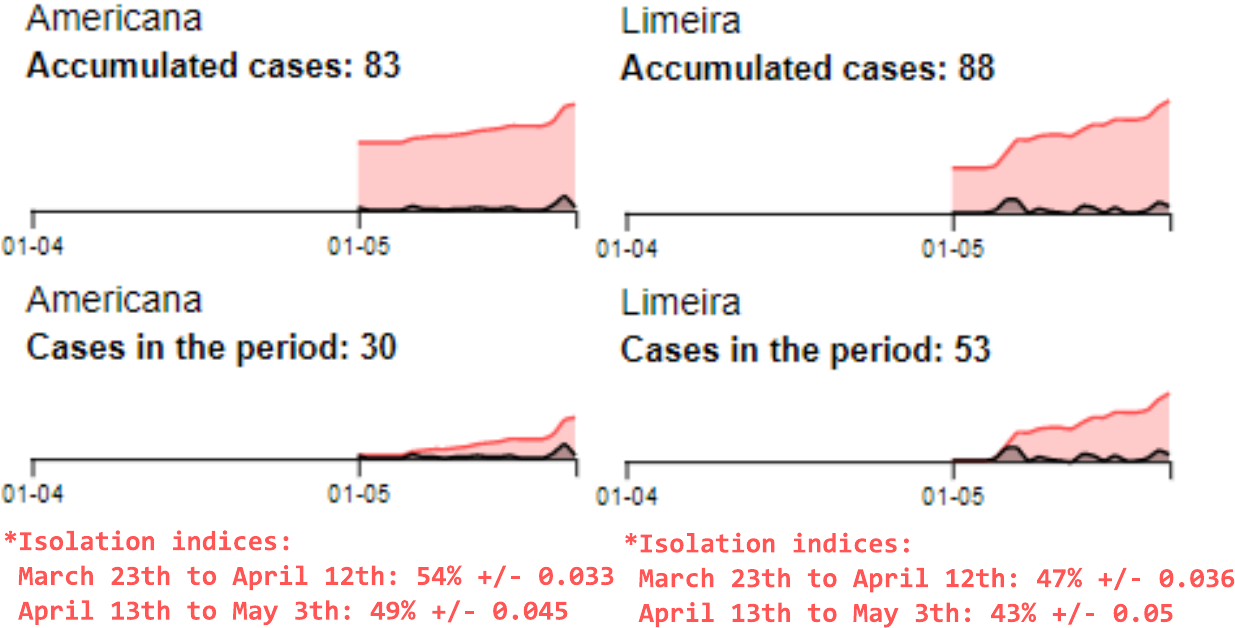}
  \caption{Dissemination curves for Americana and Limeira. In the third period of analysis, Americana shows lower dissemination, although an increase in the last few days would require attention to the next periods.}
  \label{fig:AmericanaLimeira-isolationIsBetter}
\end{figure}

Finally, the higher isolation index for both earlier periods of analysis for Americana made it possible to present a lower number of infections in this critical period, as shown in Fig.~\ref{fig:AmericanaLimeira-isolationIsBetter}.

\subsection{Analysis of bigger and regional cities}

This section aims to analyze cities closer to the São Paulo state capital (São Paulo) and other regional cities. In this case, for earlier periods, readers will see that the curves seem flattened. However, this is due to the scale used to convey the number of cases. Earlier periods are influential by the number of cases reported by the analyzed neighborhoods in this section. For instance, analysts can recall the donut charts to investigate the evolution of the number of cases visually.

\begin{figure}[!htb]
  \centering
  \includegraphics[width=\linewidth]{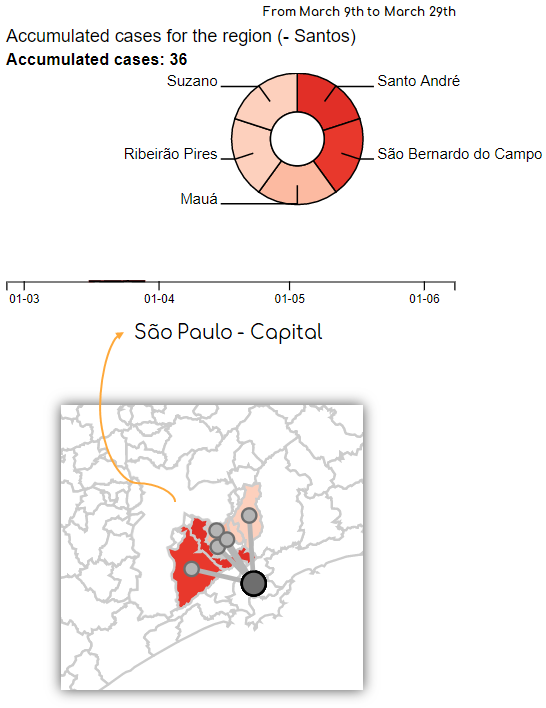}
  \caption{The situation of Santos' neighborhood. For this period, the curve seems flattened due to the graph's scale. That is, this pattern appears due to the greater number of confirmed cases.}
  \label{fig:santos-march9-march29-region}
\end{figure}

\paragraph{Santos} By March 29th, Santos's city did not present any confirmed cases. Besides that, its neighborhood was not presenting a critical situation if we look at the number of confirmed cases in Fig.~\ref{fig:santos-march9-march29-region}.

Although the situation was not very critical at such a moment, the isolation indices in Table~\ref{tab:isolation-indices-santos} could indicate difficult periods ahead. The low isolation indices are even more serious according to these cities' geolocation, which are very close to the capital, São Paulo.

\begin{table}[!h]
\centering

\begin{tabular}{lrrrr}
\toprule 
City & Isolation index           \\
\midrule 
Santos & 45\% $\pm$ 0.115 \\
Suzano & 49\% $\pm$ 0.092 \\
Ribeirão Pires & 53\% $\pm$ 0.107 \\
Mauá & 46\% $\pm$ 0.104 \\
Santo André & 46\% $\pm$ 0.123 \\
São Bernardo do Campo & 45\% $\pm$ 0.112 \\
\bottomrule 
\end{tabular}
\caption{Isolation indices in Santos' neighborhood.}
\label{tab:isolation-indices-santos}
\end{table}

\begin{figure}[!htb]
  \centering
  \includegraphics[width=\linewidth]{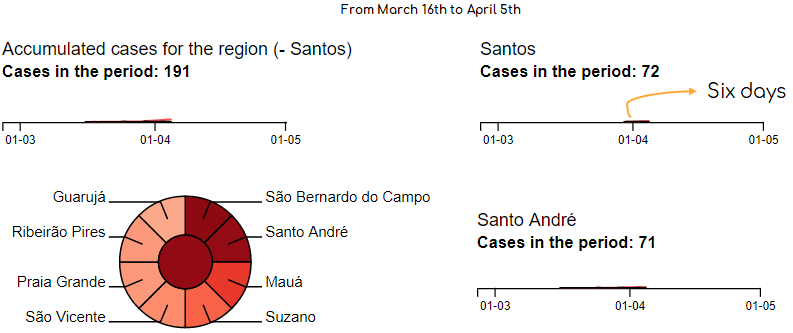}
  \caption{The situation of Santos' neighborhood from March $16$th to April $5$th. The curve of the number of cases seems flattened due to the scale -- for later periods, the number of cases is much greater.}
  \label{fig:santos-march16-april5}
\end{figure}

Moving the time window seven days further, i.e., analyzing the period from March 16th to April 5th, we can note much change in the reported number of cases. First, 191 of the 194 cases were reported in this period, in which the city of Santos reported 72 cases only in six days, as indicated in Fig~\ref{fig:santos-march16-april5}, in contrast to a greater period for Santo Andre.

\begin{figure}[!htb]
  \centering
  \includegraphics[width=\linewidth]{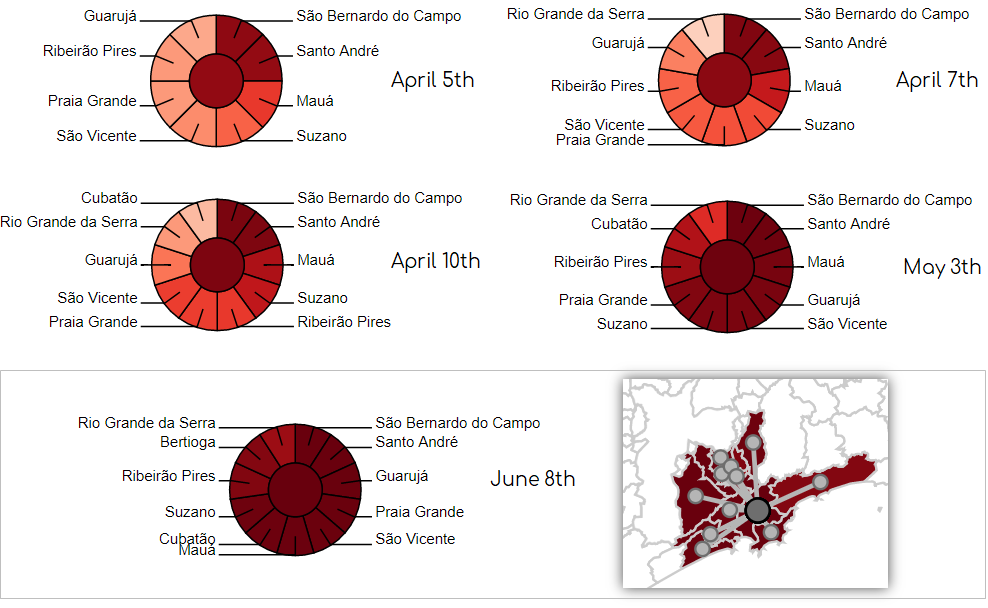}
  \caption{The situation of Santos' neighborhood on different days. There is an increasing pattern in the number of confirmed cases by inspecting the neighborhood using the color scale and donut chart representation.}
  \label{fig:santos-neighborhood-evolution}
\end{figure}

Fig.~\ref{fig:santos-neighborhood-evolution} shows the situation of Santos' neighborhood through the days. The donut chart and the color code reveal that the number of cases reported in the periods of analysis (20 days) shows an increasing pattern, configuring that such neighborhood does not reach the curve maxima. This vital information could guide decision-makers on isolation policies since it seems that the applied isolation policy up to now is not effective enough.

\begin{figure}[!htb]
  \centering
  \includegraphics[width=\linewidth]{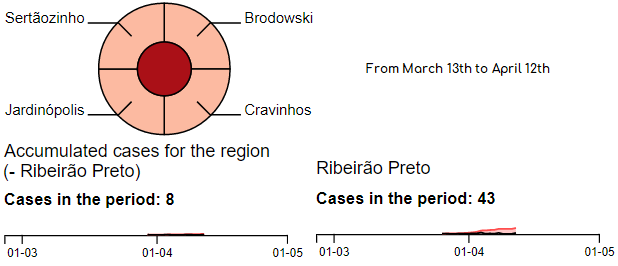}
  \caption{The situation of Ribeirão Preto's neighborhood from March 13th to April 12th. Ribeirão Preto, being the most influential city in the neighborhood, dominates the number o cases.}
  \label{fig:ribeirao-preto-march23-april12}
\end{figure}

\paragraph{Ribeirão Preto} Fig.~\ref{fig:ribeirao-preto-march23-april12} shows the number of case curves and the donut chart encoding the increase of Ribeirao Preto's neighborhood for neighborhood from March 23rd to April 12th. The visual analytics tool shows an example where the city in the analysis influences its neighborhood, i.e., while Ribeirao Preto presents 43 cases in the period, each of the other cities presents only two confirmed cases. 

The situation was different by May 3rd, in which the neighboring cities with confirmed cases jumped from four to eight. However, even with many neighboring cities with confirmed cases, Rio Preto still presents a higher number of infections than the cities in its neighborhood combined. Additionally, the increase in the number of cases seems to have a slow pace until May 3rd, which cannot be said for further periods, as we will see in the following.

\begin{figure}[!htb]
  \centering
  \includegraphics[width=\linewidth]{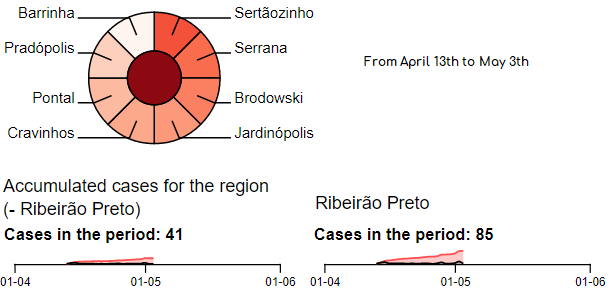}
  \caption{The situation of Ribeirão Preto's neighborhood from April 13th to May 3rd. The neighboring cities start to show an increase in the number of cases.}
  \label{fig:ribeirao-preto-april13-may3}
\end{figure}

\begin{figure}[!htb]
  \centering
  \includegraphics[width=\linewidth]{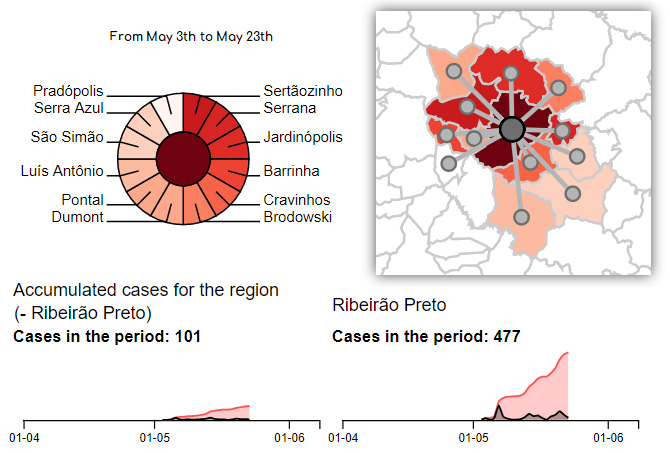}
  \caption{The situation of Ribeirão Preto's neighborhood from May 3rd to May 23rd. Besides more cities presenting confirmed cases, an increase in the number of cases can be noticed.}
  \label{fig:ribeirao-preto-may3-may23}
\end{figure}

Fig.~\ref{fig:ribeirao-preto-may3-may23} shows the situation in the neighborhood by May 23rd. Notice a big step in the number of cases reported on May 7th and an apparent increase in the number of cases reported both for the city of Ribeirao Preto and its neighborhood. Such a pattern can also be noticed by the darkening of the donut chart of the map's boundaries.

\begin{figure}[!htb]
  \centering
  \includegraphics[width=\linewidth]{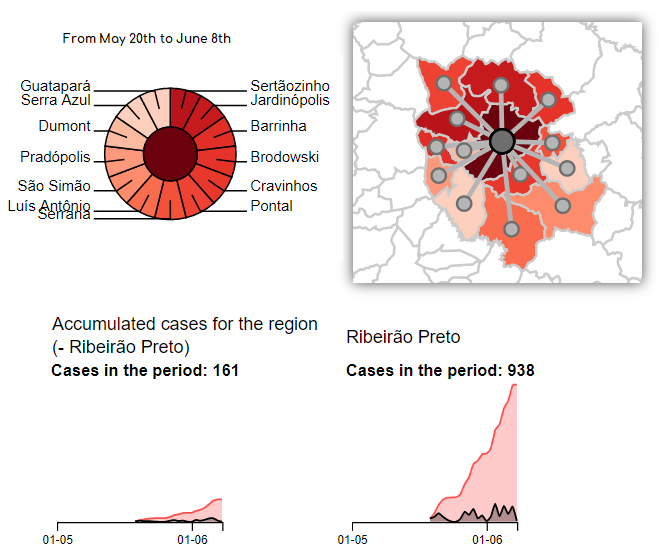}
  \caption{The situation of Ribeirão Preto's neighborhood from May 20th to June 8th. Rio Preto dominates the number of cases while other cities also show an increased pattern for their curves.}
  \label{fig:ribeirao-preto-may20-june8}
\end{figure}

Finally, Fig.~\ref{fig:ribeirao-preto-may20-june8} shows that the COVID-19 dissemination for Ribeirao Preto has not presented a decrease. Instead, the number of cases seems to be increasing rapidly, which can be dangerous due to low isolation indices presented by the cities with more critical curves – see Table~\ref{tab:isolation-ribeirao} the isolation indices for Ribeirao Preto and Sertãozinho.

\begin{table}[!h]
\centering
\begin{tabular}{lrrrr}
\toprule 
City & Isolation index           \\
\midrule 
Ribeirão Preto & 47\% $\pm$ 0.026 \\
Sertãozinho & 47\% $\pm$ 0.017\\
\bottomrule 
\end{tabular}
\caption{Isolation indices in Ribeirão Preto and Sertãozinho.}
\label{tab:isolation-ribeirao}
\end{table}

\begin{figure}[!htb]
  \centering
  \includegraphics[width=\linewidth]{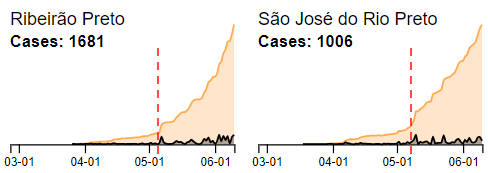}
  \caption{Although many confirmed cases, Rio Preto and São José do Rio Preto show the same rapid increase pattern, as indicated by a dashed red line.}
  \label{fig:comparacao-RibeiraoPreto-RioPreto}
\end{figure}

\paragraph{São José do Rio Preto} Fig.~\ref{fig:comparacao-RibeiraoPreto-RioPreto} shows how the number of cases' overall curve is similar between Ribeirao Preto (discussed in the previous section) and Sao José do Rio Preto are similar. In both situations, we can see a sudden increase in the number of cases indicated by a red line segment. 

\begin{figure}[!htb]
  \centering
  \includegraphics[width=\linewidth]{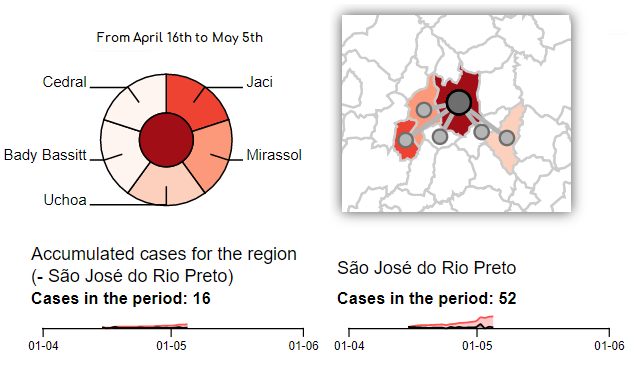}
  \caption{The situation of São José do Rio Preto's neighborhood from April 16th to May 5th. For this period, the number of cases was concentrated mainly in three cities: São José do Rio Preto, Jaci, and Mirassol.}
  \label{fig:RioPreto-april16-may5}
\end{figure}

As for Ribeirao Preto, the city of São José do Rio Preto is the most influential in its neighborhood, as shown in Fig.~\ref{fig:RioPreto-april16-may5} for the period from April 16th to May 5th. The image shows that the cases in Sao José do Rio Preto are three times greater than the accumulated cases for all the cities presenting confirmed cases. 

\begin{figure}[!htb]
  \centering
  \includegraphics[width=\linewidth]{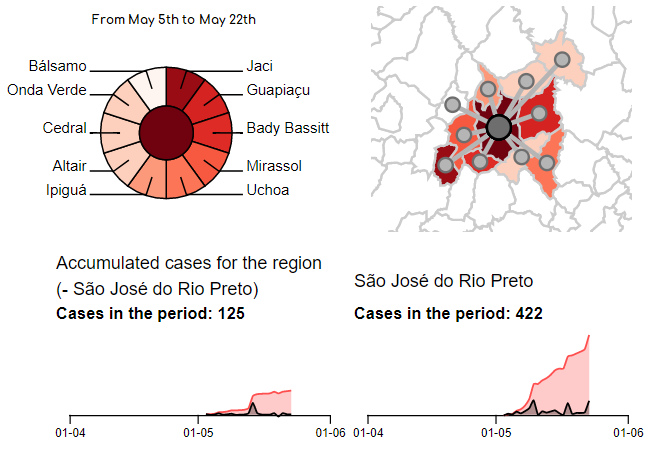}
  \caption{The situation of São José do Rio Preto's neighborhood from May 5th to May 22nd. Various cities reported confirmed cases in this period. Besides, an abnormal number of cases was reported for both cases.}
  \label{fig:RioPreto-may5-may22}
\end{figure}

Advancing the time window, Fig.~\ref{fig:RioPreto-may5-may22} shows the situation from May 5th to May 22nd. Here, besides the rapid increase in the number of cases in the following days from May 5th in São José do Rio Preto, we can note a rapid increase in the neighborhood number. While such a pattern can be explained by the interaction between the cities and consequently the dissemination of COVID-19 from São José do Rio Preto to its neighborhood - see how the increase in the neighborhood occurs later Sao José do Rio Preto -, other explanations could be accumulated the number of COVID-19 tests that were delayed and reported in such period. Fig.~\ref{fig:RioPreto-may5-may22} also suggests that the neighborhood would maintain the contamination low. However, it is not what happens in the following days.

\begin{figure}[!htb]
  \centering
  \includegraphics[width=\linewidth]{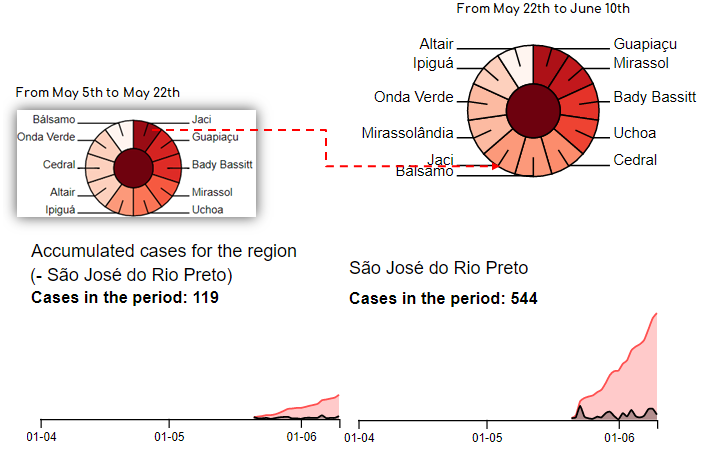}
  \caption{The situation of São José do Rio Preto's neighborhood from May 22nd to June 10th. The curves and donut charts show how the number of cases is rapidly increasing. On the other hand, the city of Jaci went from a critical situation to only confirming four cases in this period.}
  \label{fig:RioPreto-may22-jun10}
\end{figure}

Fig.~\ref{fig:RioPreto-may22-jun10} shows our last period of analysis, from May 22nd to June 10th. The curves suggest an increase in the number of cases for the city and its neighborhood, and we cannot realize any plateau in the aggregated number of cases in the neighborhood. However, it is important to emphasize how the city of Jaci went from presenting the most critical curve in the neighborhood (see the donut chart in Fig.~\ref{fig:RioPreto-may5-may22}) to only four cases in this period – such a pattern could be the result of isolation policies, but unfortunately, we do not have data to confirm. 

\begin{figure}[!htb]
  \centering
  \includegraphics[width=\linewidth]{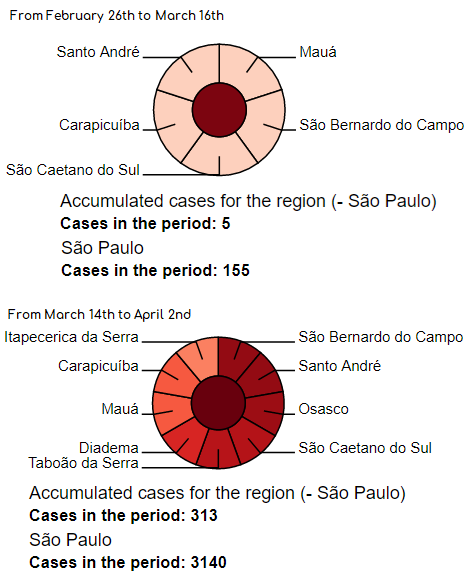}
  \caption{Comparison of São Paulo's neighborhood in the periods of February 26th to March 16th and from March 14th to April 2nd.}
  \label{fig:SaoPaulo-analysis}
\end{figure}

\paragraph{São Paulo} Finally, we analyze the evolution in the number of São Paulo cases (capital) and its neighborhood. By June 10th, São Paulo has already reported 80457 cases of COVID- 19, which is the most critical situation in the whole state (even in the whole country of Brazil). Here, we summarize the evolution of the number of cases for Sao Paulo and its neighborhood for various periods. Firstly, Fig.~\ref{fig:SaoPaulo-analysis} shows São Paulo's situation and its neighborhood from February 26th to March 16th. Note that only five cities in the neighborhood presented confirmed cases, with one case notified for each one. Advancing for the period from March 14th to April 2nd, a few more cities in the neighborhood start to present confirmed cases with the rapid increase (see Table~\ref{tab:cases-sp-neighborhood}), as shown in the donut chart.

\begin{table}[!h]
\centering
\begin{tabular}{lrrrr}
\toprule 
City & Cases in the period           \\
\midrule 
Osasco & 57 \\ 
Santo André & 70 \\
São Bernando do Campo & 70 \\
\bottomrule 
\end{tabular}
\caption{The number of cases from March 14th to April 2nd for cities in the São Paulo neighborhood.}
\label{tab:cases-sp-neighborhood}
\end{table}

\begin{figure}[!htb]
    \centering
    \subfloat[Donut chart showing a rapid increase in the number of cases.]{\includegraphics[width=\linewidth]{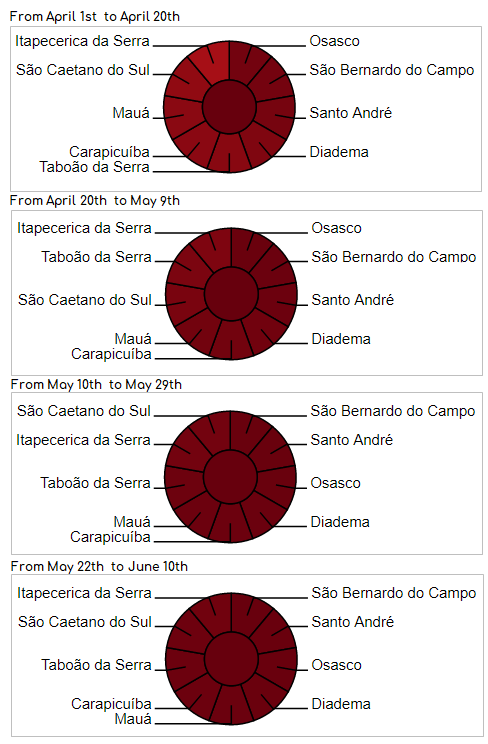}\label{fig:SaoPaulo-evolution}}    
    
    \subfloat[Curves of confirmed cases for two periods of time.]{\includegraphics[width=\linewidth]{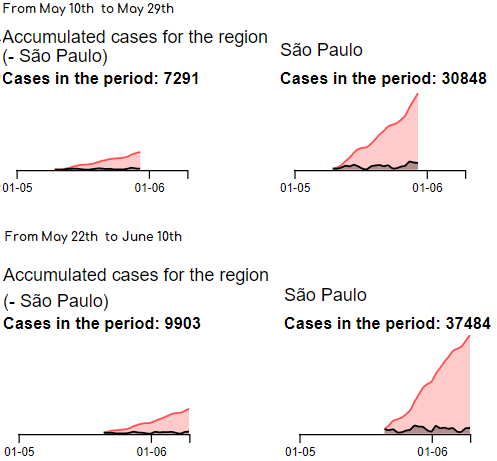}\label{fig:SaoPaulo-curves}}
    \caption{A rapidly increasing number of cases can be seen through donut charts and curves of confirmed cases in time windows.}
    \label{fig:SaoPaulo-analysis2}
\end{figure}

From this period until June 10th, São Paulo's and its neighborhood do not change in the increase in the number of reported cases. Fig.~\ref{fig:SaoPaulo-evolution} and~\ref{fig:SaoPaulo-curves} show how the dissemination of COVID-19 continues at a rapid pace in São Paulo's neighborhood.

\begin{figure}[!htb]
  \centering
  \includegraphics[width=\linewidth]{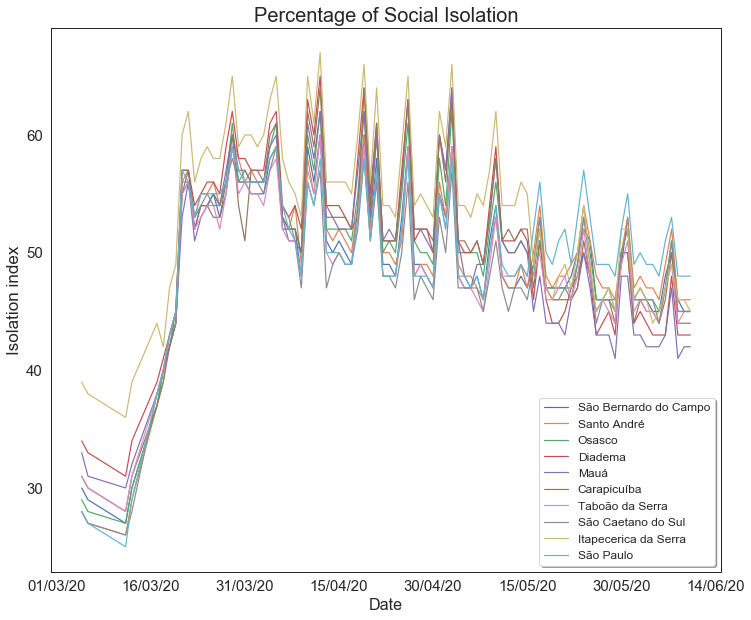}
  \caption{Social isolation of the cities in the neighborhood of São Paulo.}
  \label{fig:SaoPaulo-isolation}
\end{figure}

The cities in the donut chart of Fig.~\ref{fig:SaoPaulo-evolution} are very populous, besides interacting with themselves. On top of that, the isolation indexes are not useful, as seen in Fig.~\ref{fig:SaoPaulo-isolation}, which could aggregate even more.

\section{Discussion}
\label{sec:discussion}

Throughout the results section, we could demonstrate the usefulness of our visual analytics tool to understand the dissemination of COVID-19 in cities of interest by analyzing cities' influence on their neighborhood and vice-versa. It is important to emphasize that our tool helps analyze the evolution while it is not mainly focused on the number of cases a city or a neighborhood may present. In this case, our tool draws attention to cities or neighborhoods that present an increasing number of confirmed cases so that we believe that it could be employed even after infection by COVID-19 is controlled and employed to monitor the dissemination of other diseases. 

While we defined the neighborhood of a city as cities with spatial proximity, it is important to stress that this may not reflect the reality in some cases. For example, for a regional city, a neighborhood may be defined as the set of cities influenced by or influence cities according to some aspects, such as citizens that commute from smaller cities to greater ones to work.

\paragraph{Other indicators} Although we use the absolute case number to monitor the progression of COVID-19 dissemination, other indicators could be employed, such as a relation between the number of confirmed cases and the number of performed tests or total population. We choose the absolute number of cases in this work since it is the most straightforward approach to visualize the progression and be a well-known metric and more comfortable to understand by the general public. Notice that a more accurate picture of the epidemiological status would need a relationship between the number of tests and the neighborhood population. However, our tool offers easy interpretability and the capacity to reach a broader audience. Finally, different types of indicators (that would be useful to monitor even other diseases) can be easily incorporated in future works.

\paragraph{Visualization tool}
The visualization tool is available at RADAR\footnote{https://covid19.fct.unesp.br/analise-regional/en/}.

\section{Conclusions}
\label{sec:conclusion}

Visual analytics techniques help discover patterns that would be difficult to perceive by looking only at raw data. In this work, we employed visualization metaphors to analyze the evolution of the number of cases of COVID-19 in the São Paulo state, Brazil. Our methodology consists of visualizing the dissemination based on time windows and contrasting the number of cases in the periods of analyses with the cities' isolation indices. 

Throughout several analyses, we show how our visualization design helps analyze a city's situation according to the number of cases in a time window and its neighborhood situation. We show that our methodology emphasizes how the isolation index benefit cities regarding the dissemination, even when these cities are part of critical neighborhoods in the sense of the number of cases. 

We hope that decision-makers can use our methodology to monitor the evaluation of the number of cases in cities and neighborhoods to respond to dissemination risks quickly.

\section*{Acknowledgements}

This work was supported by Fundação de Amparo à Pesquisa do Estado de São Paulo (FAPESP) grants 18/17881-3 and 18/25755-8. We also thank the anonymous reviewers for their thoughtful suggestions on how to improve our manuscript.

\bibliography{mybibfile}

\end{document}